\documentclass[
 aip,
 amsmath,amssymb,
 reprint,%
]{revtex4-1}
\usepackage{graphicx}
\usepackage{dcolumn}
\usepackage{bm}
\usepackage{indentfirst}

\usepackage[utf8]{inputenc}
\usepackage[T1]{fontenc}
\usepackage{mathptmx}
\usepackage{etoolbox}
\usepackage{algorithm2e}
\usepackage{natbib}

\makeatletter
\def\@email#1#2{%
 \endgroup
 \patchcmd{\titleblock@produce}
  {\frontmatter@RRAPformat}
  {\frontmatter@RRAPformat{\produce@RRAP{*#1\href{mailto:#2}{#2}}}\frontmatter@RRAPformat}
  {}{}
}%
\makeatother
\begin{document}

\title[ ]{Band Structure Modeling of Perovskite Materials with Quantum ESPRESSO for Multijunction Photovoltaic Cell Optimization}
\author{Sutharsika Kumar Kalaiselvi}
 \altaffiliation[ ]{Department of Physics, North Carolina School of Science and Mathematics}

\date{\today}

\begin{abstract}
\textbf{Abstract:}
Increasing photovoltaic conversion efficiency, or PCE, has proven to be a critical factor in the transition to renewable energy. There exist strong interdependencies between the perovskite crystals and multijunction architectures within photovoltaic cell research. In present literature, there is a lack of intersection in investigating crystallographic geometry and compositional engineering with representation of computational modeling systems. In this paper, we propose a novel method for the rapid discovery of high-efficiency perovskite-based multijunction cells, specifically with silicon as the low band gap absorbing semiconductor material. We model the spatial geometries of potential perovskite candidates for high-efficiency cells using the Schrödinger Material Science Maestro, optimizing the periodic boundary conditions on the unit cell to minimize edge-bound errors. Band structure calculations using density functional theory become effective to approximate the PCE. After careful adjustments to the \verb+ibrav+ lattice parameter and parallelization, Quantum ESPRESSO was optimized for perovskite multijunction band structure calculations. Computational results on the six test-perovskite configurations demonstrate the effectiveness and superiority of our proposed representation and method, with a calculated efficiency of 
about 46\% for one of the modeled perovskites, placing it at the top of high-efficiency perovskite-Si multijunction cells. With this method, the potential exists to bring forth a new generation of photovoltaic cells that are easily manufacturable, highly efficient, and economical. 
\end{abstract}

\maketitle

\section{\label{sec:level1} Introduction}

With the demand for photovoltaic energy technologies predicted to grow significantly within the next few years and well into the future [1], a method for the discovery of new high-efficiency photovoltaic cells is essential. The efficiency of the most dominant technology in the market (i.e. Si) is 25\% in the lab and less than 20\% commercially [2]. This results in roughly 80\% of usable solar energy remaining as an untapped energy source. Concerning this, the growing demand versus the stagnant state of efficiency increase needs to be addressed. Recently, perovskites have garnered significant interest for high-efficiency applications due to being lightweight, inexpensive, and flexible while maintaining the same efficiency as non-perovskite based photovoltaic cells, among other benefits [3]. In the past few years alone, perovskites have been able to increase in efficiency from 3.8\% to over 25\%, displaying strong potential for future growth. Perovskite materials have the advantages of adjustable band gap, high absorption coefficient, long exciton diffusion length, excellent carrier mobility, and low exciton binding energy [3]. This research is based particularly on the adjustable band gap benefit that perovskites offer. This is the ability for a band gap, the energy an electron needs to move from the valence band to the conduction band, to be “tuned” to a desired amount. An example of how this would be promising for high-efficiency applications would be in a scenario where we have a semiconductor material with a hypothetical band gap of 0.6 eV, which can convert a supposed 10\% of incident solar energy into electricity, but 0.7 eV allows for the conversion of 20\% of solar energy. To be able to tune our band gap by 0.1 eV would result in an efficiency increase of 10\%. In traditional semiconductors, “tuning” the band gap is typically done by changing the doping elements, application of an external electric field, or controlled laser irradiation [4], all of which are expensive and complex processes. Meanwhile, perovskites, with the simple adjustment of the composition of elements within the crystal structure itself, would allow for such “tuning.” To utilize this unique property of perovskite crystals, another promising subject is multijunction cells.

Advances in multijunction technologies have led to the discovery of new architectures that surpass the limitations caused by current matching in series-connected cells. Multijunction cells are of interest due to the way they function. They are built using two or more semiconductor materials, hence the prefix multi, and each semiconductor converts different wavelengths of energy. With opportunity for the development of a robust system for discovering semiconductor materials, the intersection of multijunction and photovoltaic cells offers the perfect entrance for a novel method to be proposed.  

In traditional solar cell processes, silicon is the most conventional candidate as a semiconductor material. There exists a vast literature on fabrication methods for silicon in photovoltaic applications. Most of these techniques yielding in an optimized conversion efficiency utilize reactive ion etching, metal-assisted chemical etching, or strain etching [5]. All of which are high-cost operations. However, recent advancements in the fabrication protocols for silicon have improved, resulting in significant decreases in cost of production while increasing efficiency. In a multijunction application, especially as the first work considering the intersection of these fields, it is beneficial to model perovskites that are in an architecture with silicon. Although their efficiency is low, they offer a significant foundation on which to build this method so that eventually perovskite-perovskite structures can be modeled.  

Advancements in PCE have been halted due to factors such as the cost of fabrication during trial phases, often resulting in semiconductors being unfit for the job. Computational approaches have been promising in this aspect, as they can model numerous amounts of crystals without spending nearly as much money, albeit with computational costs. These computational calculations can approximate the band structure of a given crystal system, which can then provide valuable information on the predicted PCE of the crystal semiconductor in a PV cell. Existing research using computational approaches mainly uses density functional theory, or DFT, to approximate the band structure of crystal structures. Issues with density functional theory arise when approximating complex crystal configurations such as perovskites. For band structure approximations, Quantum ESPRESSO or the opEn-Source Package for Research in Electronic Structure, Simulation, and Optimization, is most used [6]. Quantum ESPRESSO utilizes the plane-wave basis set and crystallographic information to approximate the interactions of electrons within a crystal system by integration along the Brillouin Zone. This computational system is not suitable for modeling various crystals in a series, specifically ones with complex structures, and minor changes occurring with the replacement of elements when “tuning” for a specific band gap. Further, limitations with the system make it so that modeling multijunction cells is nearly impossible, requiring the need for an alternative method. 

Recently, there have been advancements in the modeling of complex crystals using Quantum ESPRESSO. These methods focus on optimizing the choice of high-symmetry points within the crystal lattice [7], yet they are unable to model crystals as complex as perovskites in an efficient manner. The way Quantum ESPRESSO is built using DFT is the primary reason for this issue. The core of the code in Quantum ESPRESSO is a quantum engine whose main purpose is to perform Hamiltonian builds, i.e., the application of the Hamiltonian operator to molecular or Bloch orbitals and related operations, and to solve the quantum-mechanical equations that determine them and their response to external perturbations [7]. Moreover, current methods to optimize this process for complex crystal structures lack functionality in a multijunction application, along with the elemental idea of compositional engineering, a topic integral to this research. Hence, we propose adjustments to the \verb+ibrav+ parameter, pertaining to the information regarding crystal lattice structure, as well as parallelization to streamline the calculations process for large-scale modeling of perovskite crystals.  

Although it may be evident that a direct relationship is present among the topics, namely, multijunction cells, perovskites, and computational modeling, interestingly, there is a lack of research at the intersection of these fields. 

Our contributions to this opportunistic subject can be summarized as follows: (1) We present a solution to overcome the current barrier faced between the demand and innovation in semiconductor materials for photovoltaic cell applications. (2) We introduce a computational method for the modeling of complex crystal structures, namely, optimizing Quantum ESPRESSO for this application. (3) We present the first-ever work that brings together two promising fields of photovoltaic efficiency in a manner that is easily reproducible and consistent with the goals of improving PCE.   

\section{\label{sec:level1} Related Work}

In this section, we review the most related works regarding representation of perovskite and multijunction cells, specifically done using computational methods.

\subsection{\label{sec:level2} Perovskite Structure}

Perovskite crystal structures are defined by their $ABX_3$ stoichiometry. A and B are cations which reside in the corner and the body center of the pseudo-cubic unit cell and X is the anion which occupies the face center [8]. Perovskite materials have a direct tunable bandgap, i.e. by substituting iodine with bromine in the X-site results in a band gap increase from 1.55 eV to roughly 2.3 eV, among other methods that involve changes to the A and B sites of the structure as well. Crystals with these structure are of great use in optoelectronic applications due to long charge-carrier lifetimes, excellent charge-carrier mobilities, and high defect tolerances [4]. For the purposes of using Quantum ESPRESSO, a cubic perovskite structure is considered most optimal due to the method of integration along the Brillouin Zone (BZ). Perovskite thin films can be reduced to 100 nm in thickness, while more common semiconductor materials used in photovoltaic cells are typically 50-60 $\mu$m [5], all of these put together make it an ideal candidate for high efficiency PV cells.

\begin{table}[h!]
 \caption{\label{tab:table1} Lattice coordinates for the Pm3m space group of which vast majority of artificially fabricated perovskites fall under. This is especially useful for retaining the structure during DFT approximations [9].\\}

\centering
 \begin{tabular}{||c c c||} 
 \hline
 Site & Location & Coordinates \\ [0.5ex] 
 \hline\hline
 A & (2a) & (0,0,0) \\ 
 B & (2a) & ($\frac{1}{2}$, $\frac{1}{2}$, $\frac{1}{2}$) \\
 X & (6b) & ($\frac{1}{2}$, $\frac{1}{2}$, 0)($\frac{1}{2}$, 0, $\frac{1}{2}$)(0, $\frac{1}{2}$, $\frac{1}{2}$)\\ [1ex] 
 \hline
 \end{tabular}
\end{table}

\begin{figure}
    \centering
    \includegraphics[scale=0.45]{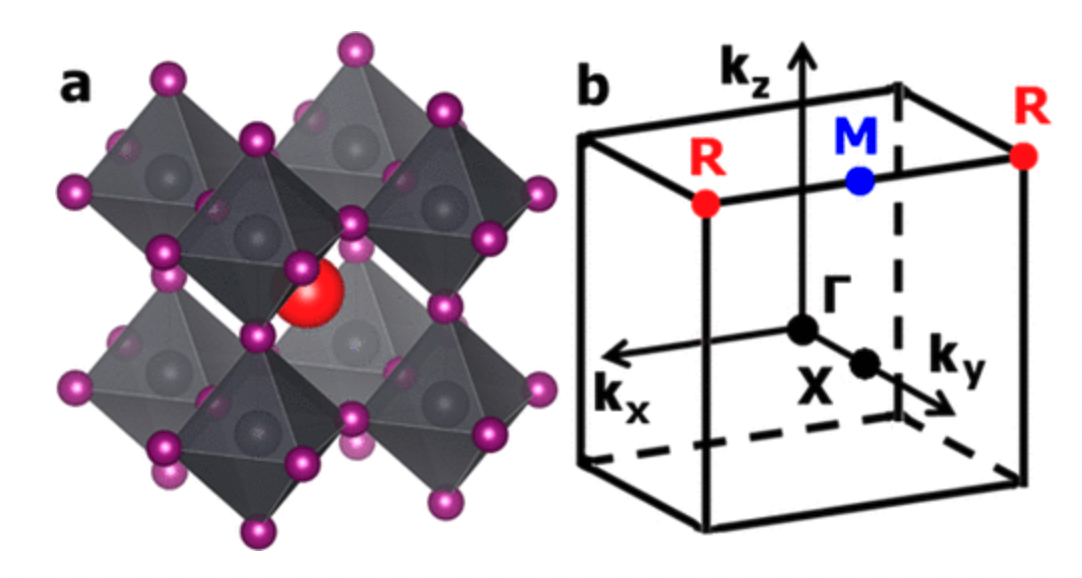}
    \caption{(a) Real space 3D view of the Pm-3m reference cubic crystal structure of metal halide AIPs or HOPs of general formula $ABX_3$.(b) Reciprocal space 3D view showing the first BZ of the Pm-3m space group. Points of high symmetry in the cubic BZ are indicated by conventional letters; $\Gamma$ denotes the origin of the BZ, X is the center of a square face at the BZ boundary, M is a center of a cube edge, and R are vertices of the cube [10].}
    \label{fig:enter-label}
\end{figure}

\begin{figure}
    \centering
    \includegraphics[scale=0.35]{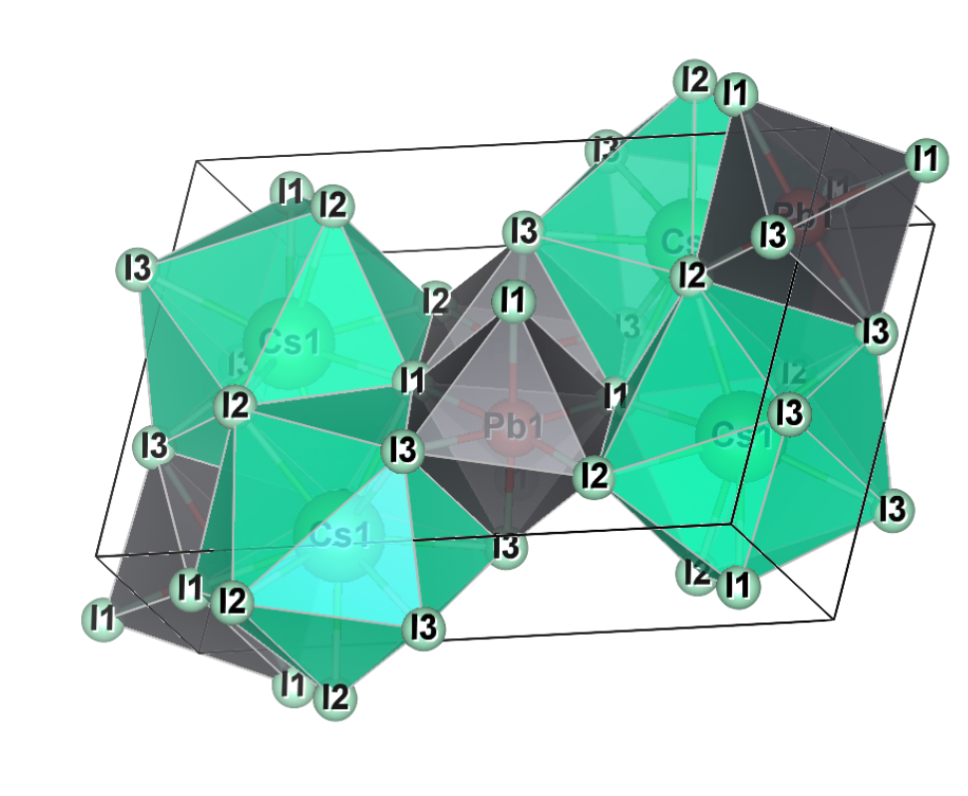}
    \caption{Optimized $CsPbI_3$ in the Pm3m space group, Cs is a cation occupying the A-site, Pb is an anion occupying the B-site, and I is a halide ion occupying the X-site.}
    \label{fig:enter-label}
\end{figure}

\subsection{\label{sec:level2} Multijunction Architecture}
Multijunction cells are made of several layers of different semiconductor materials. The radiation that passes through the first layer is absorbed by the latter layers, resulting in the absorption of more light per unit area [11]. This makes multijunction solar cells more efficient than single junction solar cells,  view \textit{figure 3}.

\begin{figure}
    \centering
    \includegraphics[scale=0.55]{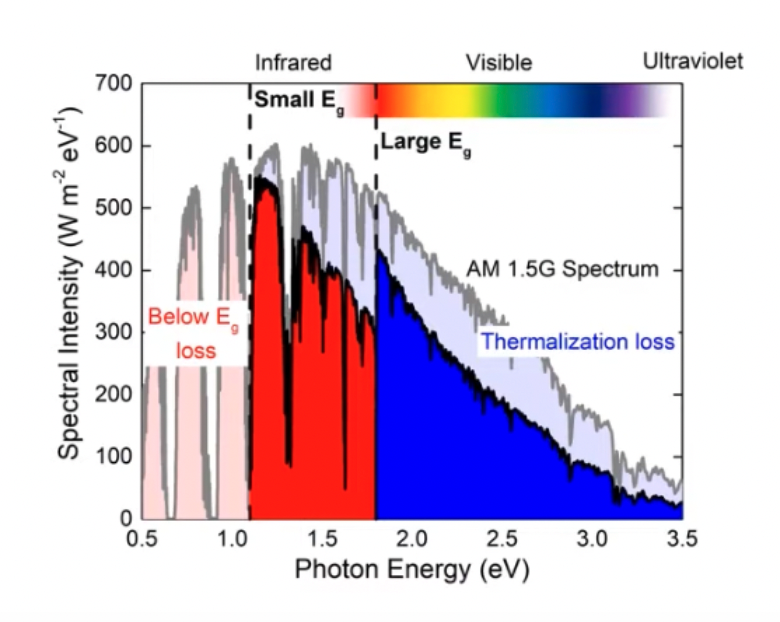}
    \caption{Representation of the different wavelengths of energy that are absorbed with two semiconductors in a multijunction architecture, avoiding below band gap and thermalization losses [31]}
    \label{fig:enter-label}
\end{figure}

Only two semiconductor materials are of use because cost rapidly increases with the addition of each new semiconductor material within the architecture, resulting in a decreased efficiency vs. cost ratio. In our experiment, one of these is set to be silicon (Si). This is in part because of the extensive establishments already refined towards the fabrication of silicon semiconductors as they are the most standard commercial PV cell semiconductor. In order to transition to a fully perovskite-perovskite multijunction cell, this intermediate step is essential. Silicon has a fixed and well-known band gap, which also helps with multijunction PCE calculations. 

There are three representative categories of two-junction solar cells: two-terminal (2T) monolithically integrated, four-terminal (4T) mechanically stacked, and 4T optically coupled multijunction cells [12]. This research focuses on 4-T multijunction cell architectures, see figure 4. The two-terminal monolithically integrated cell, although promotes direct recombination and therefore higher efficiencies compared to other architectures, presents major limitations with its application for perovskite crystals. Since the current flowing through the series-integrated is governed by Kirchhoff's law, bandgap and thickness of each subcell should be precisely managed to achieve photo-current match at the maximum power point. Additionally, the bottom cell serves as the substrate for the growth of the top cell, the successful deposition of the top cell depends entirely on the quality of the bottom cell. The charge recombination layer must also have low electrical resistance as well as good transparency in the near-infrared region to guarantee that the long wavelength photons can reach the bottom cell [12]. 

Optically coupled two-junction cells are inefficient due to their scalability issues. With the way the architecture fits within the multijunction framework, increasing this in size for commercial or industrial applications becomes an expensive endeavor [11], leaving 4T mechanically stacked (b) in figure 4 as the best candidate architecture. The component cells are electrically independent due to the physical separation of the two subcells [12]. The total PCE is a sum of the two subcells. This is extremely convenient for computational applications, as you can independently calculate the PCE of each subcell and add it together to calculate total PCE, simplifying the PCE calculation and analysis processes. 

\begin{figure}
    \centering
    \includegraphics[scale=0.2]{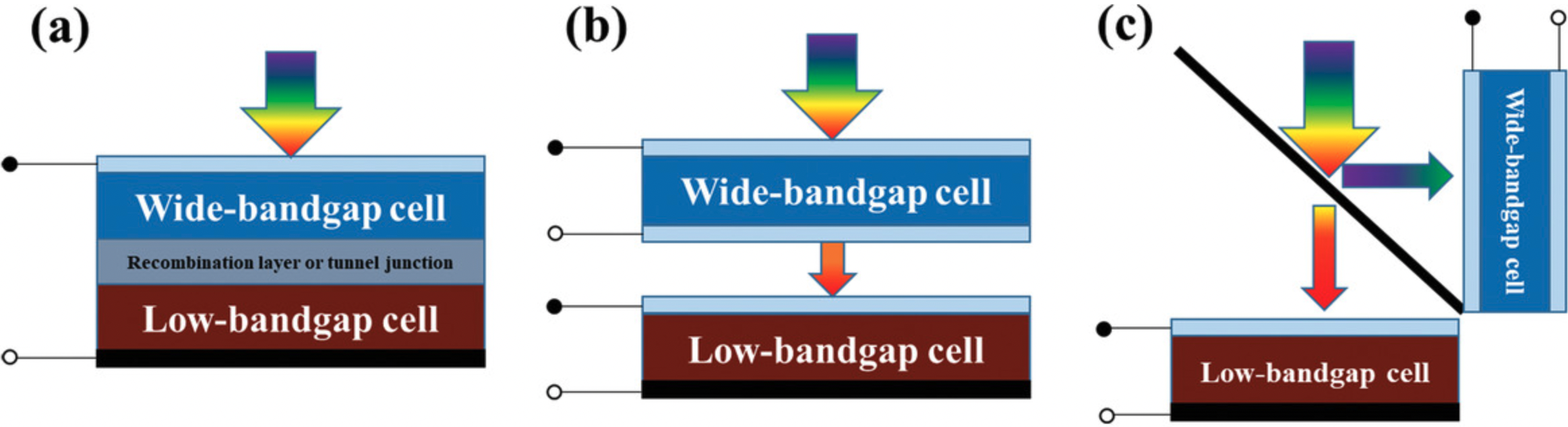}
    \caption{(a) represents the 2T monolithically integrated cell type, (b) 4T mechanically stacked, and (c) 4T optically coupled [12].}
    \label{fig:enter-label}
\end{figure}

\subsection{\label{sec:level2} Quantum ESPRESSO}

Quantum ESPRESSO, or the Quantum opEn-Source Package for Research in Electronic Structure, Simulation, and Optimization, as the name suggests, is an open-source package built upon Density Functional Theory (DFT) to simulate the interactions of atoms in a unit cell to calculate band structure as well as other electronic properties [13].

We calculate the electronic structure of real materials and their physical properties by the \textit{ab-initio} method, particularly electrons, governed by the laws of quantum mechanics [14].

Equation (1) is the wave-function, and calculating the wave-function of the electron in the potential field (V) is calculated by solving the Schrödinger equation (2). 

\begin{eqnarray}
\lambda + p = h
\\
\left\{%
\frac{-h}{2m}\nabla^2 \Psi(\textbf{r},t) + V(\textbf{r},t) = ih\frac{\partial\Psi(\textbf{r},t)}{\partial t}%
\right\}%
\label{eq:one}
\end{eqnarray}

"To reduce complexity, we remove time (t) as one of the parameters the function is based off of" [14]. For many simulations, potential field is rarely a function of time due to the simulation not being largely affected by it. This means the spatial and temporal parts of the wave function can be separated, resulting in equation (3). 


\begin{eqnarray}
\varepsilon \Psi(r) = \Psi(\textbf{r})
\left[
\begin{array}{c}
\frac{-h^2\nabla^2}{2m} + v(\textbf{r})
\end{array}\right]\;
\\
E\Psi(r_1, r_2, ..., r_N) = \Psi(r_1, r_2, ..., r_n)
\left[
\begin{array}{c}
\frac{-h}{2m}\sum_{i=1}^{N} \nabla^2_i + 
\\
\sum_{i=1}^{N} V(r_i) + ...
\end{array}\right]
\end{eqnarray}

"The calculation gets significantly more complex when solving the Schrödinger equation as a real physical system consisting of a large n number of atoms" [14]. The Schrödinger equation becomes a coupled many-body system [7], represented by equation (4). Current high performance computing is also limited to only performing $10^{18}$ floating-point operations (flops), while solving the wave-function would require much higher computational power [13].

DFT functions are primarily based on the Hohenberg-Kohn Theorems and the equations governing this. The \textbf{Hohenberg-Kohn Theorem 1} states that the ground state density $n(\textbf{r})$ determines the external potential energy $v(\textbf{r})$ within a trivial additive constant. This means that once a wave function is solved, the other properties, such as density can also be determined. The theorem also states the reverse, for a given density, the potential can be uniquely determined.

Kohn-Sham also states that the nonlinear eigenvalue functions created to determine density of a system need self-consistent solutions, this requires the wave-function to be expanded into a basis set. 

Quantum ESPRESSO utilizes the plane-wave basis set in particular, along with pseudopotential values to minimize periodic boundary condition errors and simplify the calculation [14]. 
\\
\\
\textit{More information regarding the theory behind Quantum ESPRESSO can be found in source [13].}

\section{Methods}

To develop a new method for the rapid-discovery of perovskite crystals for multijunction photovoltaic cell applications, a process of optimizing the crystals must be created. For this, it was decided upon to test $TiO_2$, a common material used as a buffer or transition layer in solar cells. Using the Schrödinger Material Science Maestro Suite, we were able to model the structure and optimize the unit cell for Quantum ESPRESSO. 

\begin{figure}
    \centering
    \includegraphics[scale=0.25]{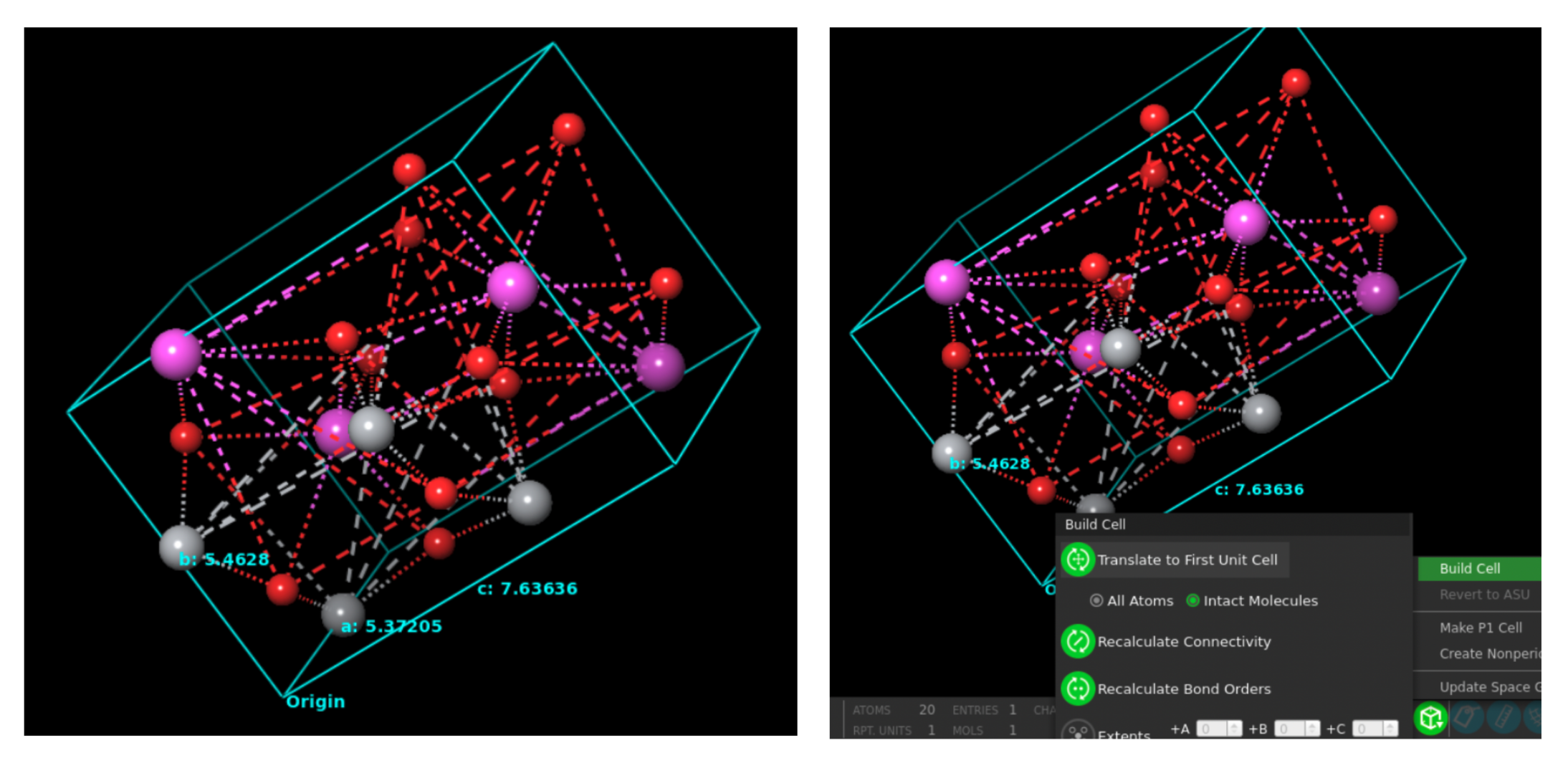}
    \caption{The following steps have been taken on $TiO_2$ to optimize the structure: Translate First Unit Cell (Intact Molecules), Recalculate Connectivity, and Recalculate Bond Orders.}
    \label{fig:enter-label}
\end{figure}

\begin{figure}
    \centering
    \includegraphics[scale=0.35]{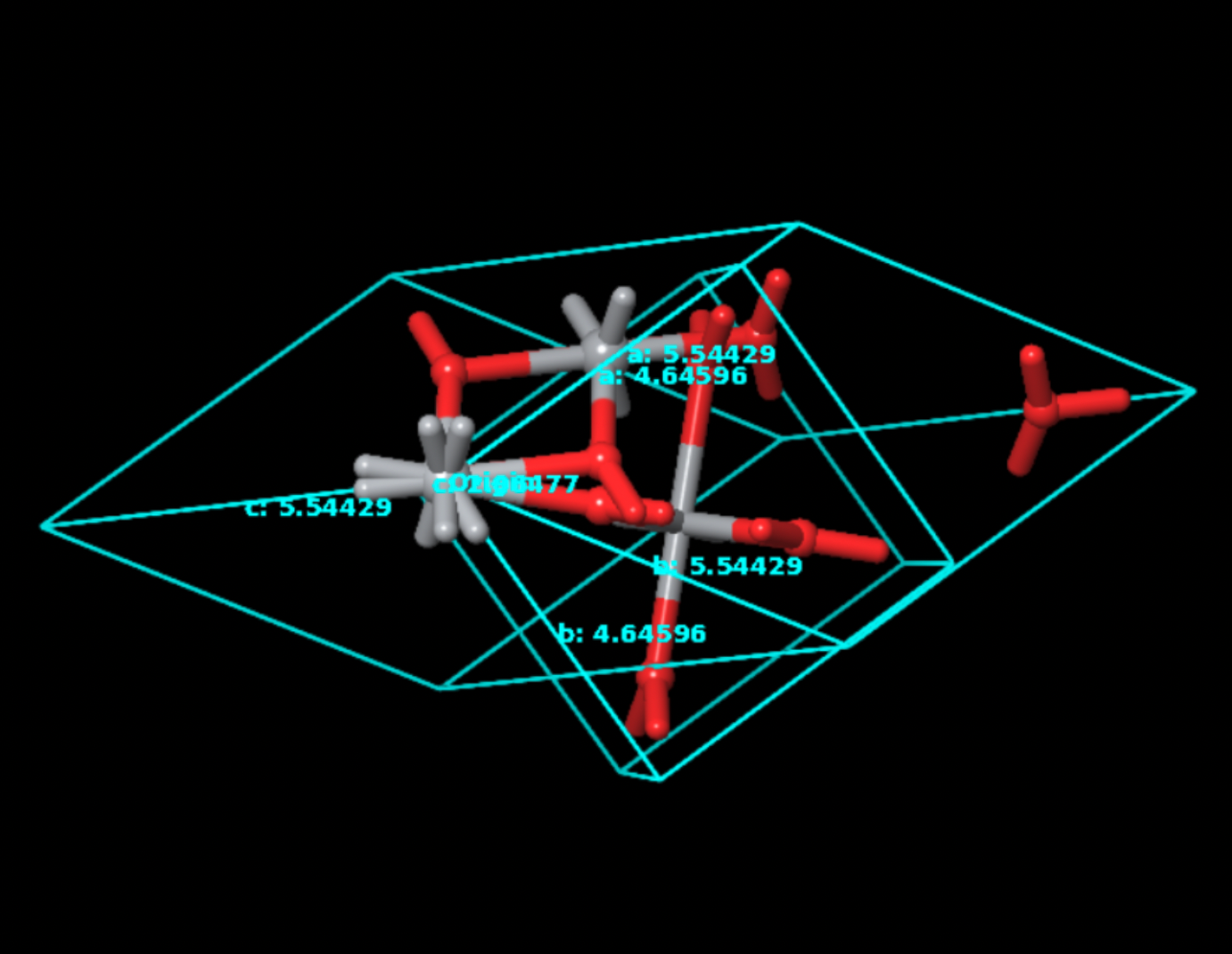}
    \caption{The result of the actions taken in figure 5 result in an optimized $TiO_2$ until cell.}
    \label{fig:enter-label}
\end{figure}

From this, we were able to configure an input file to execute the following required Quantum ESPRESSO calculations for band structure. 
\\
\\
1) \verb+scf+ - self-consistent field calculations
\\
2) \verb+nscf+ - non-self consistent field calculations
\\
3) \verb+bands.x+ - band calculations (band gap is determined from this step, and is the difference in energy levels from the conduction band and valence band. 
\\
4) \verb+post-processing+ - for a plottable format.
\\

The Local Density Approximation, equation 5 (LDA) and the Gradient Generalized Approximation, equation 6 (GGA) were tested on this input file, resulting in two band structures, and therefore two band gaps. $TiO_2$ was also in-part selected as a preliminary test due to the experimental band gap being well-known for this structure. The experimental band gap for $TiO_2$ is 3.0 eV [15], while the LDA resulted in an approximated 2.18 eV, and GGA in a 2.7 eV. This shows that GGA is likely a better approximation method compared to LDA due to it being more physically consistent. Additionally, the true exchange-correlation functional of DFT should depend on the gradient of the density [16], hence Gradient Generalized Approximations would be more accurate.

\begin{eqnarray}
E_{xc} = \int n(r)\epsilon_{xc}(n(r))dr
\\
E_{xc} = \int n(r)E_{GGA}(n(r),|\nabla{n}(r)|)dr
\label{eq:one}
\end{eqnarray}

After completing the $TiO_2$ calculation and creating the input file for $CaTiO_3$, a few problems arose. These were regarding the \verb+ibrav+ parameter. A perovskite crystal has a unique cubic structure that is not defined by the \verb+ibrav+ parameter, which defines the points of high symmetry along the BZ upon which to perform integration.

\begin{figure}
    \centering
    \includegraphics[scale=0.35]{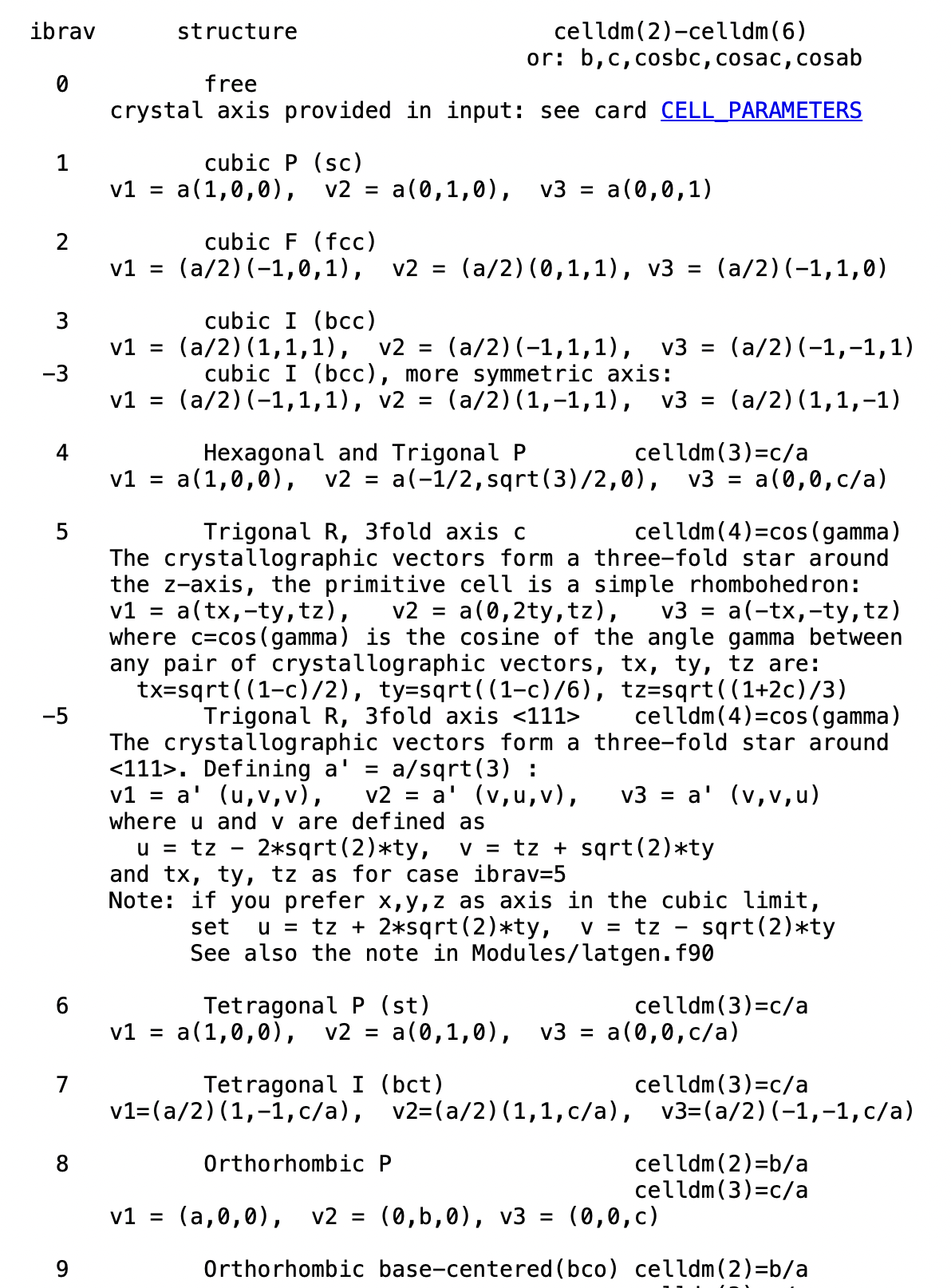}
    \caption{Quantum ESPRESSO documentation pertaining to the ibrav parameter. 1,2,3,-3 refer to cubic configurations, of which, none are optimized for perovskite crystal modeling [13].}
    \label{fig:enter-label}
\end{figure}

In order to optimize the \verb+ibrav+ parameter, a new parameter needs to be included following the perovskite crystal coordinates we discussed in the related works. Since the coordinates always remain the same, it is possible to create a new parameter within the input file that can take into account these specific cell dimensions. With this input parameter, it is possible to replace the crystal elements composing the perovskite structure without having to update or adjust the lattice parameter every time compositional engineering is implemented. This reduces the complexity of the system, while also maintaining computational accuracy.

On a note back to using the GGA for calculating the band structure, there is an error associated with this calculation, and the way we optimized QE for perovskite modeling, results in a relatively consistent error. This can be accounted for by taking out the error from the computational results, which effectively gives us the experimental band gap of the system that is being modeled.

Another issue that was encountered was parallelization. The Pittsburgh supercomputer, in which all calculations were run, allows for parallel runs, but the Quantum ESPRESSO parallelization was not optimized for this purpose. we attempted to address this issue by using two terminal windows, but attained minimal success, this is still something we hope to improve for the future. Once all key issues regarding the computational framework had been addressed it was time to execute the computational modeling protocol. 

The k-resolved Density of States is useful to understand the band structure and determine accuracy. If the k-resolved DOS, since it is resolved to momentum should be similar to the band structure calculations, which can be seen with a comparison of figure 8 to figure 9, meaning the silicon band structure is likely accurate. This is useful for maintaining computational viability. 

\subsection{Perovskite Selection}

The perovskites modeled were selected through an intensive literature review, and later composition engineering. In present literature, it is widely agreed upon that MAPBI3 and FAPBI3 are the most promising candidates for high-efficiency applications. 

With these two base perovskites, we proceeded to look at the atomic configuration and determined that replacing the A-site with Cesium may yield positive results, I then also noticed that replacing Pb with Sn, tin in the B-site might also produce promising results. These perovskites' band gap were computationally approximated. 

\begin{figure}
    \centering
    \includegraphics[scale=0.35]{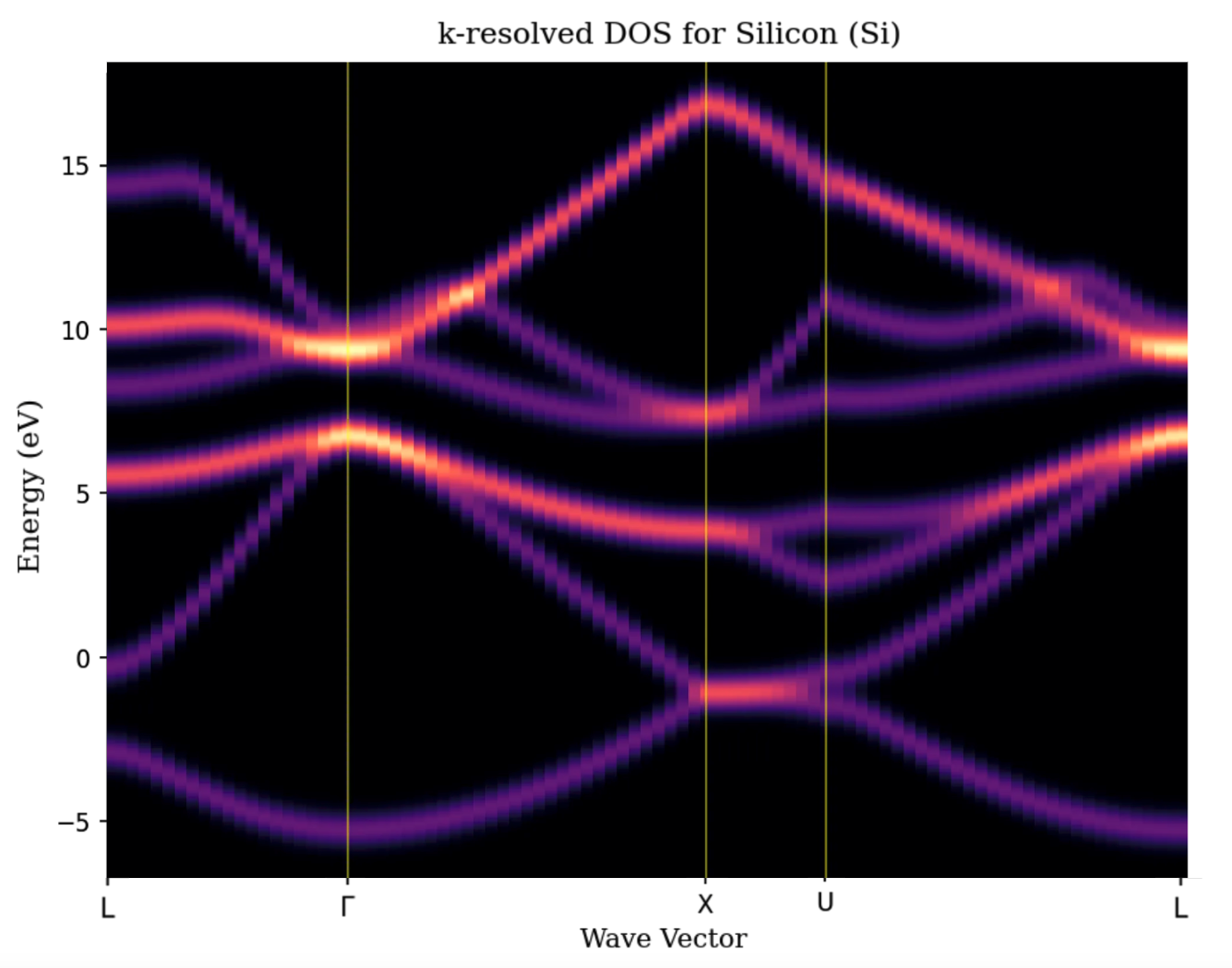}
    \caption{k-resolved Density of States for Silicon (Si), resolved to k, momentum along the Brillouin Zone, showing the number of electronic states per unit energy, plottable code optained from [14].}
    \label{fig:enter-label}
\end{figure}

\section{Results}

\begin{figure}
    \centering
    \includegraphics[scale=0.35]{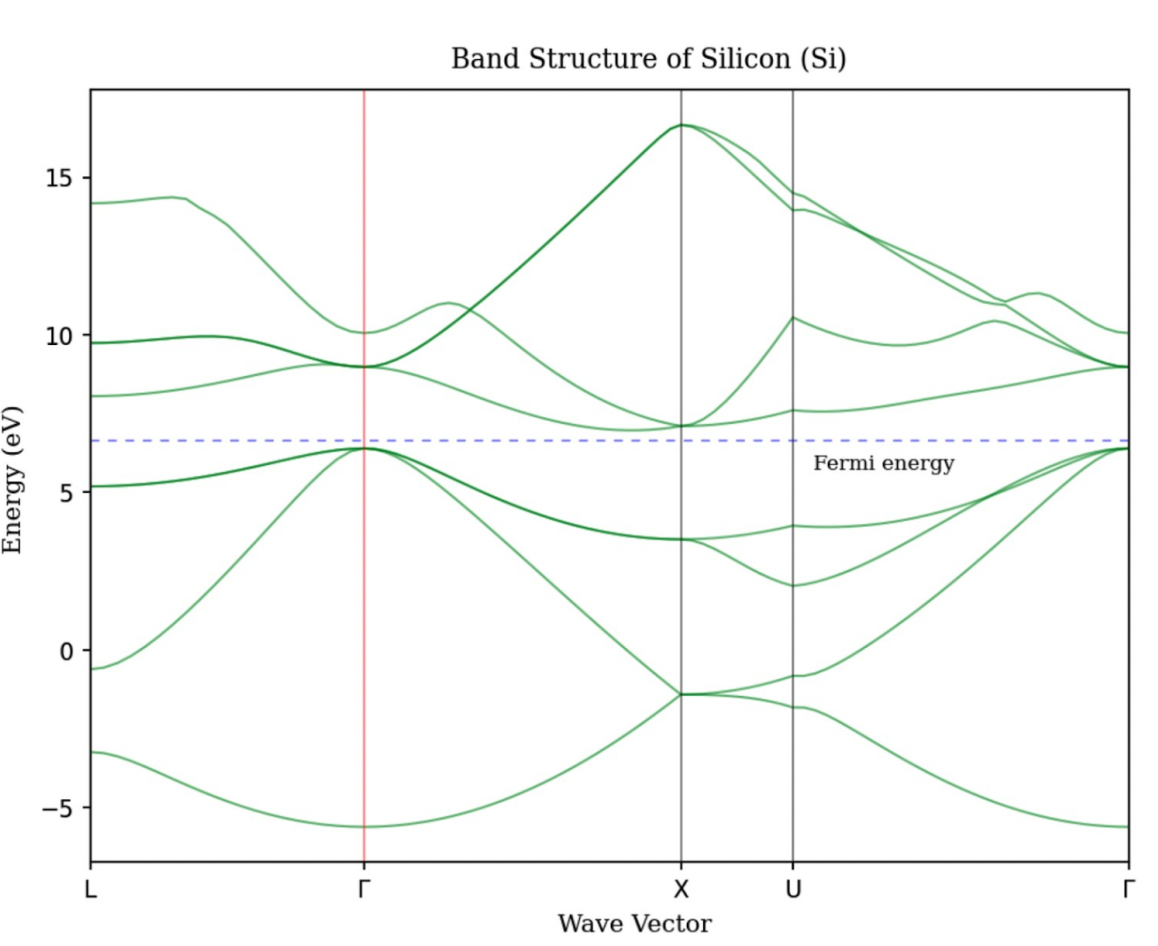}
    \caption{Silicon band structure calculated using GGA approximation in QE, indirect band-gap semiconductor.}
    \label{fig:enter-label}
\end{figure}

\begin{figure}
    \centering
    \includegraphics[scale=0.35]{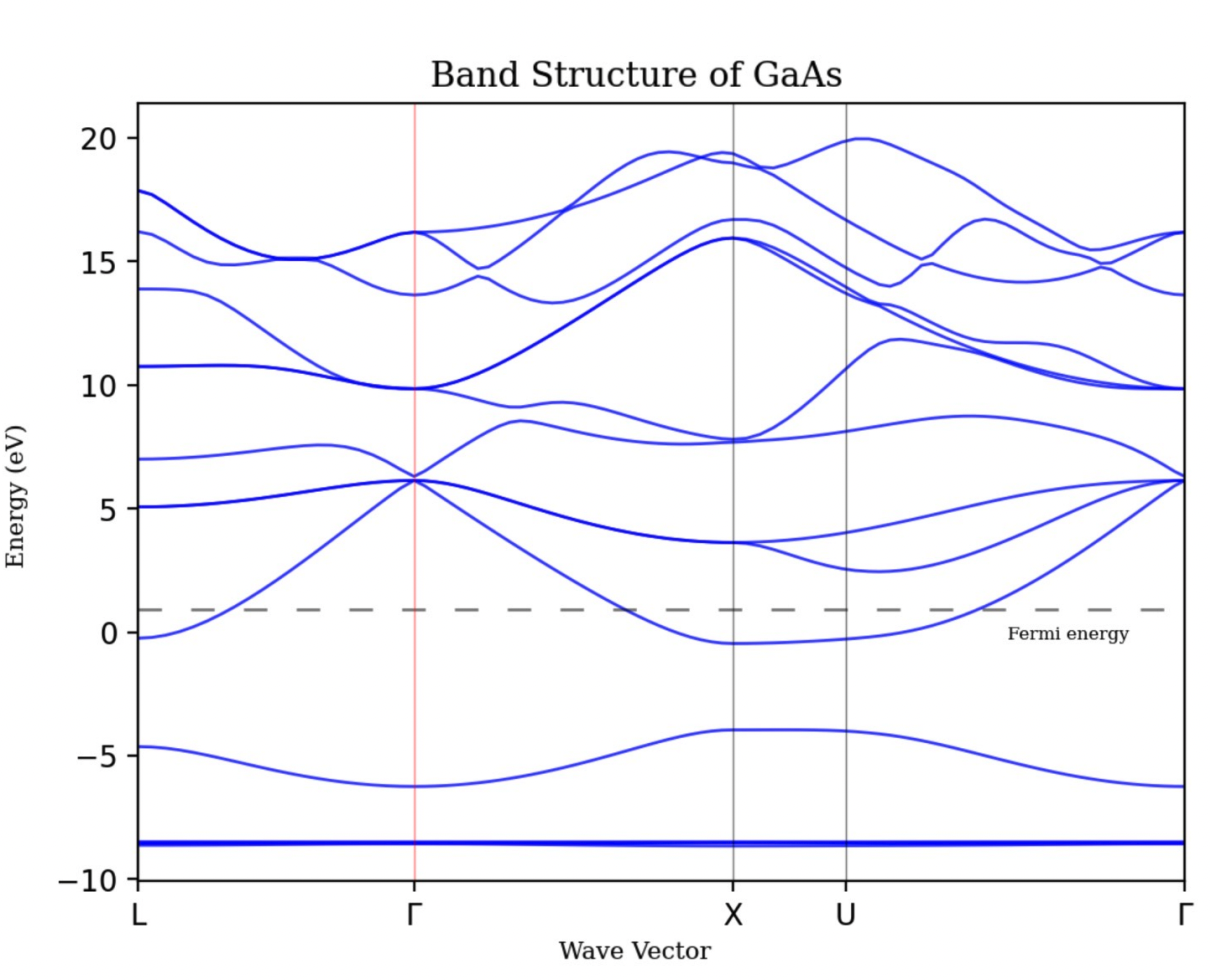}
    \caption{GaAs band structure calculated using GGA approximation in QE, direct band-gap semiconductor, code obtained from [14].}
    \label{fig:enter-label}
\end{figure}

Table 2 shows the results that were obtained during this research, of which we see that 3 are promising high efficiency candidates. Additionally figure 9 and 10 are band structures of proposed bottom cells, silicon and GaAs. In this paper, we primarily focus on silicon due to its established manufacturing processes as a stepping stone to later perovskite-perovskite cells. However, it would not be detrimental to create a dataset including the PCE of GaAs and perovskite semiconductors in a multijunction architecture. 

\begin{table*}
\caption{\label{tab:table3} Results of the modeled crystal structures using the novel method developed within this research.}
\begin{ruledtabular}
\begin{tabular}{cccc}
\textbf{Structure} & \textbf{Approximation Method} & \textbf{Computational Band Gap} & \textbf{Space Group}\\
 Si & GGA & 0.73 eV & Fd3m \\
 GaAs & GGA & 0.58 eV & F43m \\
 CaTiO3 & GGA & 2.12 eV & Pnma\\
 MaPbI3 & GGA & 1.1 eV & Pm3m\\
 FAPbI3 & GGA & 1.02 eV & Pm3m\\
 CsPbI3 & GGA & 1.3 eV & Pnma\\
 MASnI3 & GGA & 1.2 eV & Pm3m\\
 FASnI3 & GGA & 1 eV & Pm3m\\
 CsSnI3 & GGA & 1 eV & Pnma\\
\end{tabular}
\end{ruledtabular}
\end{table*}

\section{Analysis}
Using the experimental PCE of these crystals, we were able to account for computational error during the approximations and determine this final heatmap of PCE for dual-junction 4T mechanically stacked PV cells (figure 11). After doing this, we can now plot the estimated efficiencies of these crystals, resulting in CsPbI3-Si being the strongest dual-junction silicon based high-efficiency cell. 

\begin{figure}
    \centering
    \includegraphics[scale=0.25]{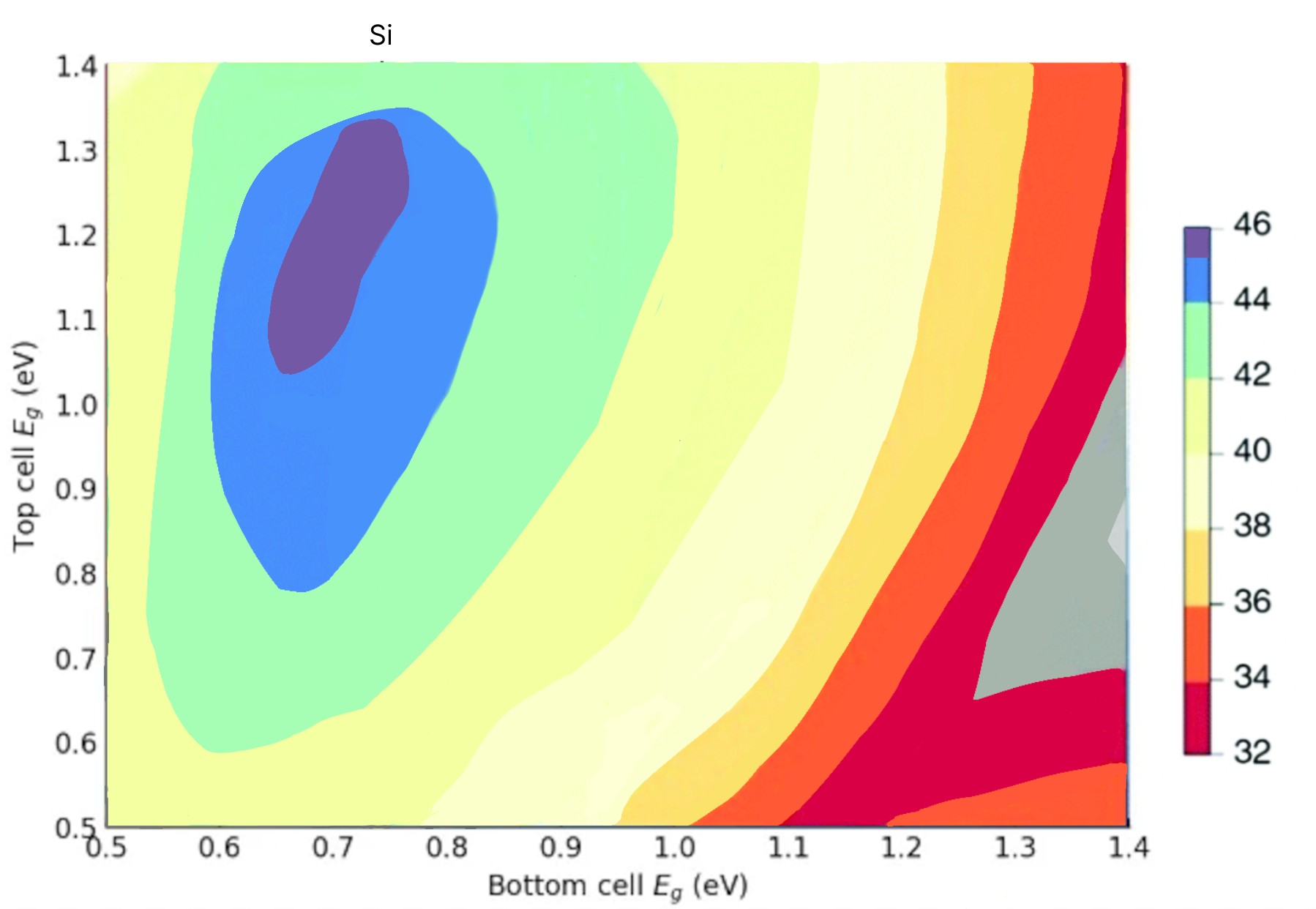}
    \caption{Heat map of PCE conversion efficiencies, showing CsPbI3-Si is in the most optimum zone with a PCE of about 46\% [adapted from 20].}
    \label{fig:enter-label}
\end{figure}

\section{Conclusions \& Future Work}
Previously there was no method that was able to overcome the high-cost low-reward situation faced by high-efficiency solar cell discovery. By working in the intersection of perovskite, multijunction, and computational physics research, the first ever method/work tackling this issue has been proposed. 

We were able to overcome the limitations that arose with modeling a complex many-body system perovskite in Quantum ESPRESSO by using a novel approach on the lattice ibrav parameter. Additionally, of the six perovskites that were modeled in a multijunction architecture with silicon, 3 resulted in being promising high-efficiency candidates. CsPbI3-Si had an estimated computational PCE of 46\%, with the current highest efficiency solar cell at 47.6\% as a Gallium based triple-junction cell. Approaching this record-high efficiency in just 6 models, and with an economical process.

Future work could include increasing the size of a perovskite-perovskite and perovskite-Si multijunction PCE dataset and open-source it for the purpose of being utilized by the general public. Additionally, a validation process can be established by fabricating the high-efficiency candidate perovskites to attain an experimental backbone to computational results. With promising results, additional research could provide great steps forward in the race against climate change.

\begin{acknowledgments}
I extend my gratitude towards my mentors, Dr. Bennett, and Dr. Falvo from the North Carolina School of Science and Mathematics, Department of Physics, for providing the resources and guidance necessary to complete this project. We would also like to thank the Pittsburgh High Performance Computing Center for the access to a supercomputer, without which, this research would not have been possible. 
\end{acknowledgments}

\end{document}